\documentclass[12pt]{article}
\usepackage{subfigure}
\usepackage{epsfig}
\usepackage{epsf}
\begin{document}
\begin{center}

 {\Large \bf INFLUENCE OF WETTING PROPERTIES ON DIFFUSION IN A CONFINED FLUID}

\vskip 3cm

{\Large Yannick Alm\'eras (*), Jean-Louis  Barrat ($\dagger$), Lyd\'eric Bocquet  (*,$\dagger$)}
\end{center}
\vskip 3cm

(*)  
Laboratoire de Physique
ENS-Lyon and CNRS, 69364 Lyon Cedex 07, France

\bigskip

($\dagger$) D\'epartement de Physique des Mat\'eriaux, Universit\'e Claude-Bernard
Lyon-I and CNRS, 69622 Villeurbanne C\'edex, France

\vskip 7cm

{\it Proceedings of the workshop "Dynamics in confinment",
Institut Laue-Langevin, Grenoble, January 2000.
(http://www.ill.fr/Events/confit.html)}

\newpage

\vskip 3cm

\centerline{\bf Yannick Alm\'eras , Jean-Louis  Barrat, Lyd\'eric Bocquet }
\vskip 3cm

\begin{abstract}
We briefly discuss how the wetting properties of a 
fluid/solid interface can indirectly influence the diffusion
properties of fluid confined between two solid
walls. This influence is related to the variability
of the hydrodynamic boundary conditions at the interface,
which correlates to the wetting properties.
\end{abstract}

\section{The boundary condition at a liquid-solid interface}

At a macroscopic level, it is well known that the relative velocity of 
a fluid with respect to the solid vanishes at a liquid-solid 
interface.  
This is the  "no-slip" boundary condition, which although very general
does not have a microscopic justification. At a microscopic level, 
it is however necessary to take into account a possible "slip"
of the liquid on the solid surface. The amount of slip
is quantified by introducing a slipping length in the boundary 
condition at the solid interface, which in general reads
\begin{equation}
{\partial v_t  \over \partial z}|_{z=z_{w}} = -
 {1 \over \delta} v_t|_{z=z_{w}}  ,
\label{def_slip}
\end{equation}
where $v_t$ is the tangential velocity at the boundary,
and $z_w$ is the position of the boundary (which is assumed here to be a plane
perpendicular to the $z$ axis).
The "slip" length $\delta$ which appears in this equation can
be interpreted as the length one has to extrapolate the velocity
field of the fluid into the solid to obtain a vanishing 
value.  Equation (\ref{def_slip}) can also be interpreted as expressing
 the continuity of the  stress (or momentum flux) at the boundary.
At the boundary, the viscous stress $\eta{\partial v_t  \over \partial z}$ in the fluid
 is then equal to a fluid friction stress between the solid
and the liquid, $\kappa v_t$. $\delta$ is the ratio of the viscosity $\eta$ to the friction constant
$\kappa$. The usual "no-slip"
boundary condition corresponds
to $\kappa=\infty$, while at a free boundary $\kappa=0$.

In previous work \cite{TR90,BB94} it was shown
that even for surfaces which  are smooth at the atomic scale,
slip is usually a small effect. A small (atomic) corrugation of the wall
is enough to produce a "no-slip" boundary condition. This accounts for the findings of experiments performed with the surface force apparatus
\cite{georges}. 
However,  this result appears to break down when the
solid surface is strongly nonwetting for the liquid, i.e. 
when a liquid drop on this solid substrate has a large contact angle.
In that case, it appears \cite{BB99} that the slip length $\delta$ can become
much larger than the molecular size. Physically, this can be traced back
to the fact that the liquid does not "want" to be in contact with the solid. 
Hence  a microscopically thin depletion layer forms  between
the bulk fluid and the solid, making the momentum transfer 
much less efficient and effectively decoupling the fluid from the substrate.
In the following we discuss how the diffusion of a molecule will be 
affected when a thin liquid film is  confined between two identical 
 parallel 
plates that
are characterized by a "partial slip" boundary condition
such as (\ref{def_slip}). The quantity we focus on
is the relative change of the diffusion constant parallel
to the plates as a function of the distance $h$ between the plates, 
\begin{equation}
\Delta  = {D_\parallel(h) -D_{bulk} \over D_{bulk}}.
\end{equation}
We will be interested in cases where the confinment is moderate (typically $h$
is larger than 10 molecular sizes), so that the film is 
still in a clearly fluid state.

\section{Confinement effects on diffusion.}

\subsection{Qualitative discussion.}

In this section, we briefly describe, at a qualitative level, 
how the boundary conditions can influence the diffusion of a 
molecule confined in a pore. Two complementary points of view are possible,
and yield essentially identical results \cite{epl}. The first one is 
a microscopic, "mode-coupling", type of approach. The idea is the following.
Very generally, the diffusion constant of a tagged particle 
can be written as
\begin{equation}
D= \int_0^\infty dt <\vec{v}(t).\vec{v}(0)> 
\end{equation}
In the bulk, two contributions to the velocity autocorrelation function
$<\vec{v}(t).\vec{v}(0)>$ of the tagged particle
can be isolated \cite{epl}. A short time part describes the "rattling" motion
in the cage formed by the neighbours. A more subtle contribution,
which appears for long times, is related to the so called "backflow" 
effect. The idea is that the initial momentum of the 
particle is transferred at intermediate times to the long wavelength, hydrodynamic motion of the fluid. According to the Stokes equations,
this momentum diffuses away from the tagged particle. However, 
the properties of the diffusion equation imply that
a fraction of this momentum eventually returns to the origin and "pushes"
the tagged particle. Let us now consider how this mechanism is modified
by confinment. First of all, the "rattling" contribution is not
expected to change, since it is governed by the local environment.
The hydrodynamic backflow, on the other hand, will be strongly modified.
If the confining boundaries correspond to a "no-slip" situation,
they will absorb the incoming momentum. In that case the amount of backflow 
will be reduced, and  $\Delta$ will be negative. On the other hand
in a case of perfect slip the momentum will be reflected at the boundary, and the backflow effect will be enhanced. This argument can be made quantitative
in both cases \cite{epl} and has in fact been used to intrepret experimental 
results on free standing liquid crystal films \cite{lc}. However, it turns
out that a quantitative calculation for the case of partial slip is difficult.

An alternative, more macroscopic line of thought consists in computing
the {\it mobility} $\mu_\parallel(z)$ of a particle
at a distance $z$ from a solid wall, with a boundary condition
\ref{def_slip},  using macroscopic
hydrodynamics. The diffusion constant in a fluid slab is then obtained
using the Einstein relation between diffusion and mobility, averaged
over the thickness $h$ of the slab. For the no-slip or perfect slip cases,
such a calculation was shown to yield results identical to 
those obtained within the mode coupling approach. For the general
"partial slip" boundary conditions, it offers the advantage of being
tractable analytically. The method and results are summarized 
in the next section.

\subsection{Hydrodynamic estimate of the diffusion constant.}

  Consider a spherical
particle of diameter $R$ moving past
a solid boundary characterized by equation \ref{def_slip},
with a constant velocity parallel to the boundary. The 
mobility is obtained by calculating the viscous drag on the particle, which
implies solving the Stokes equation for the velocity and pressure 
fields. This  can be achieved using the 
method of reflections \cite{hb}. In this method,
the    velocity field at  zeroth order corresponds to 
the one obtained for an infinite fluid, and
therefore obeys the correct boundary condition on the particle.
A  first correction term is introduced to obtain the 
correct boundary condition on the wall, therefore violating the
no-slip boundary condition on the particle. A third correction
is introduced to correct again the boundary condition on the particle.
Assuming convergence of the series, one eventually ends up with
a velocity field that has the correct behaviour both at the particle
surface and at the solid boundary. The force 
can then be calculated from the pressure tensor on the particle surface.
The details of the calculation are described in \cite{stage}. 
Here we only quote the result, which gives the force on a particle
at an altitude $z$ from the wall, with velocity $\vec{U}$ as
\begin{equation}
\vec{F}=\frac{-6\,\pi\,\eta\,R\,\vec{U}}{1-\frac{9}{16}\,\frac{R}{z}\:{C}
\!\left[\frac{\delta}{z}\right]}
\label{eq:funmur}
\end{equation}
where ${\rm C}\!\left[\frac{1}{y}\right]=-\,\frac{1}{6}\,y^2-\frac{1}{2}\,y-\frac{2}{3}+\left(\
\frac{1}{6}\,y^3+\frac{2}{3}\,y^2+\frac{2}{3}\,y\right){\rm E}(y)+\frac{8}{3}\,y\:{\rm E}\!\left(\frac{y}{2}\right)$ and 
$E(y)= e^y {\rm Ei}(1,y)$, with ${\rm Ei}(1,y)$ the exponential integral
function. The altitude dependent mobility is then averaged over the 
chanel to compute the effective diffusion constant.
When applied to the extreme cases of no-slip
or perfect slip, this formula yields 
a relative decrease or increase, respectively, of the diffusion constant, in accordance with the qualitative analysis made in the previous section.
The results for $\Delta$ as a function of $h/R$ and $\delta/R$ are summarized 
in figure \ref{3d}. An increase in the diffusion constant can be observed as
soon as $\delta $ becomes larger than the pore size.
\begin{center}
\begin{picture}(0,0)
\epsfig{file=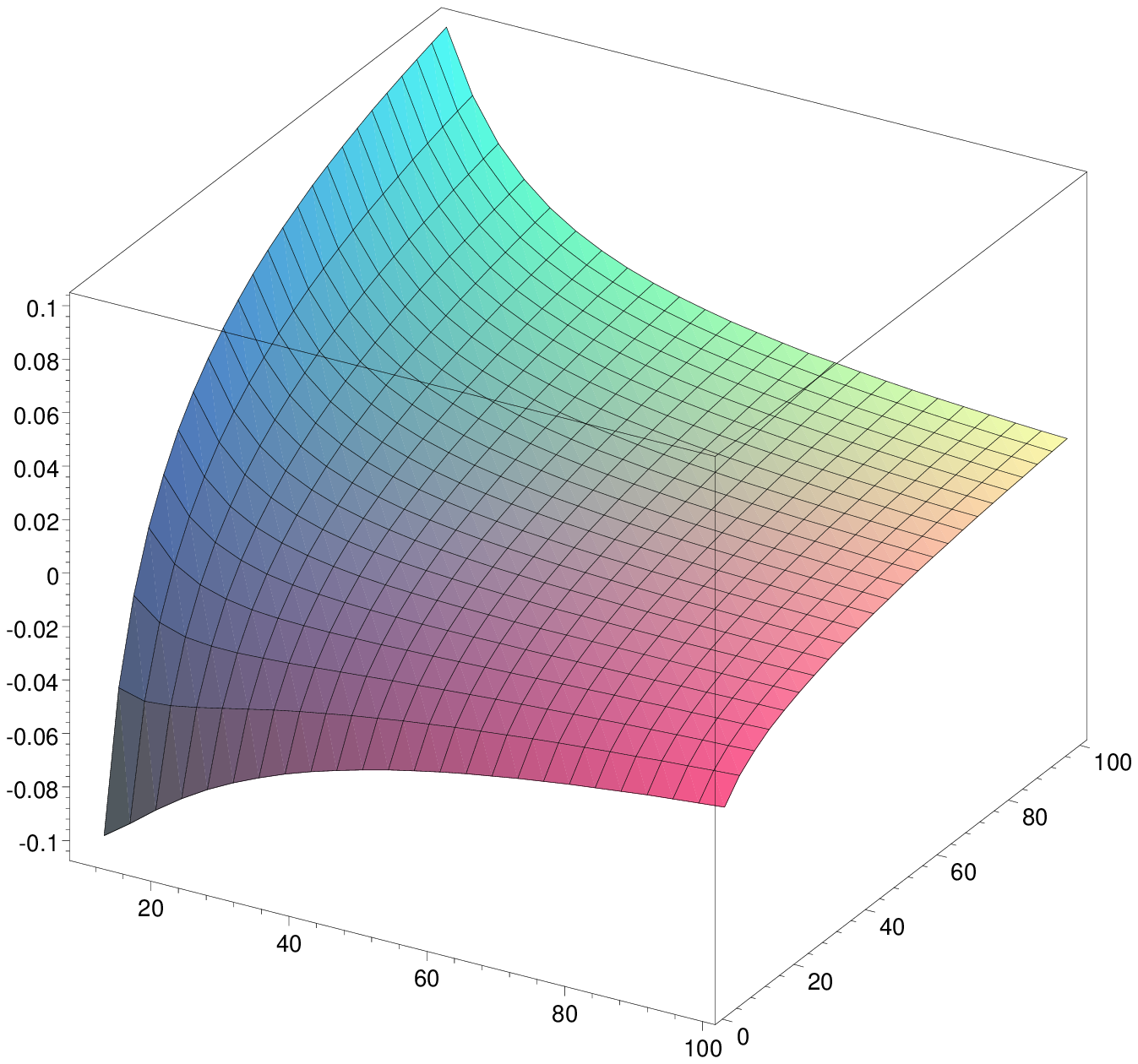,width=13.5cm,height=5.5cm}%
\end{picture}%
\setlength{\unitlength}{4736sp}%
\begingroup\makeatletter\ifx\SetFigFont\undefined%
\gdef\SetFigFont#1#2#3#4#5{%
  \reset@font\fontsize{#1}{#2pt}%
  \fontfamily{#3}\fontseries{#4}\fontshape{#5}%
  \selectfont}%
\fi\endgroup%
\begin{picture}(3000,4500)(4501,-6373)
\put(4401,-4546){\makebox(0,0)[lb]{\smash{\SetFigFont{17}{20.4}{\familydefault}{
\mddefault}{\updefault}$\frac{\Delta D}{D}$}}}
\put(7351,-5600){\makebox(0,0)[lb]{\smash{\SetFigFont{17}{20.4}{\familydefault}{
\mddefault}{\updefault}$\frac{\delta}{R}$}}}
\put(5526,-5901){\makebox(0,0)[lb]{\smash{\SetFigFont{17}{20.4}{\familydefault}{
\mddefault}{\updefault}$\frac{h}{R}$}}}
\end{picture}
\vskip-1.5cm
\begin{figure}[h]
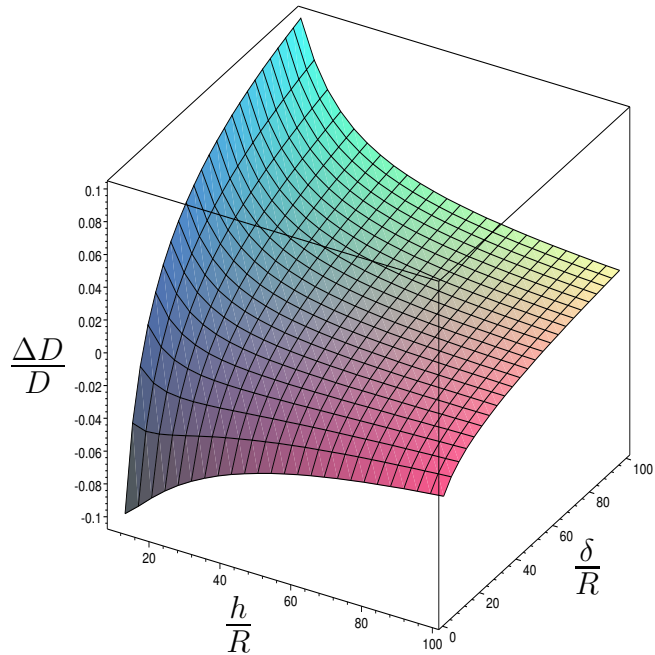

\caption{\textit{Reduced diffusion coefficient, $\Delta=\frac{\Delta D}{D}$ 
versus $\frac{h}{R}$ and $\frac{\delta}{R}$}}
\label{3d}
\end{figure}
\end{center}

\subsection{Molecular dynamics simulations.}

\begin{figure}[h]
\centering
\mbox{\subfigure[$\delta^\star=1,7$]{\epsfig{figure=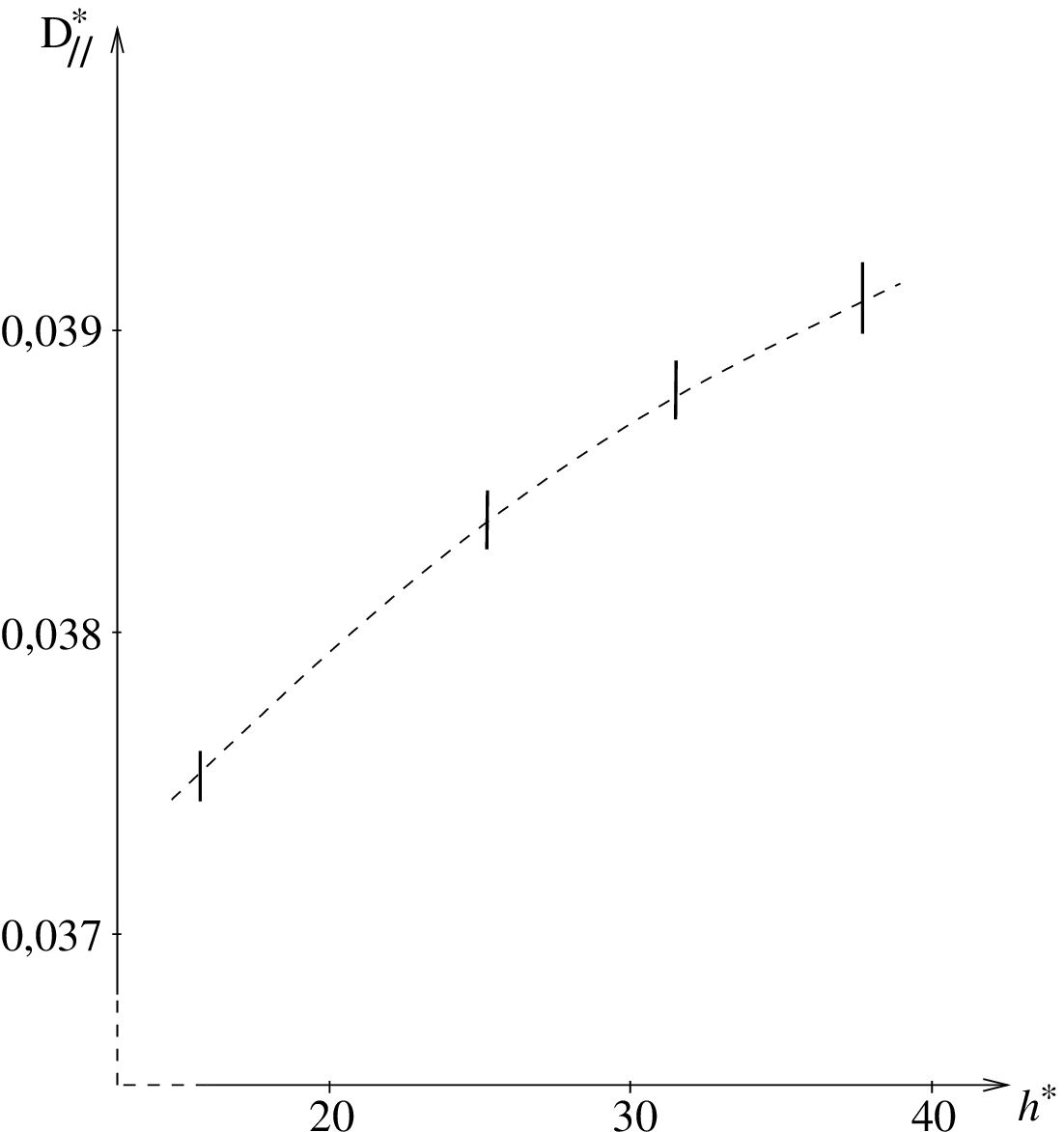,width=7cm}}\qquad
\qquad\subfigure[$\delta^\star=41$]{\epsfig{figure=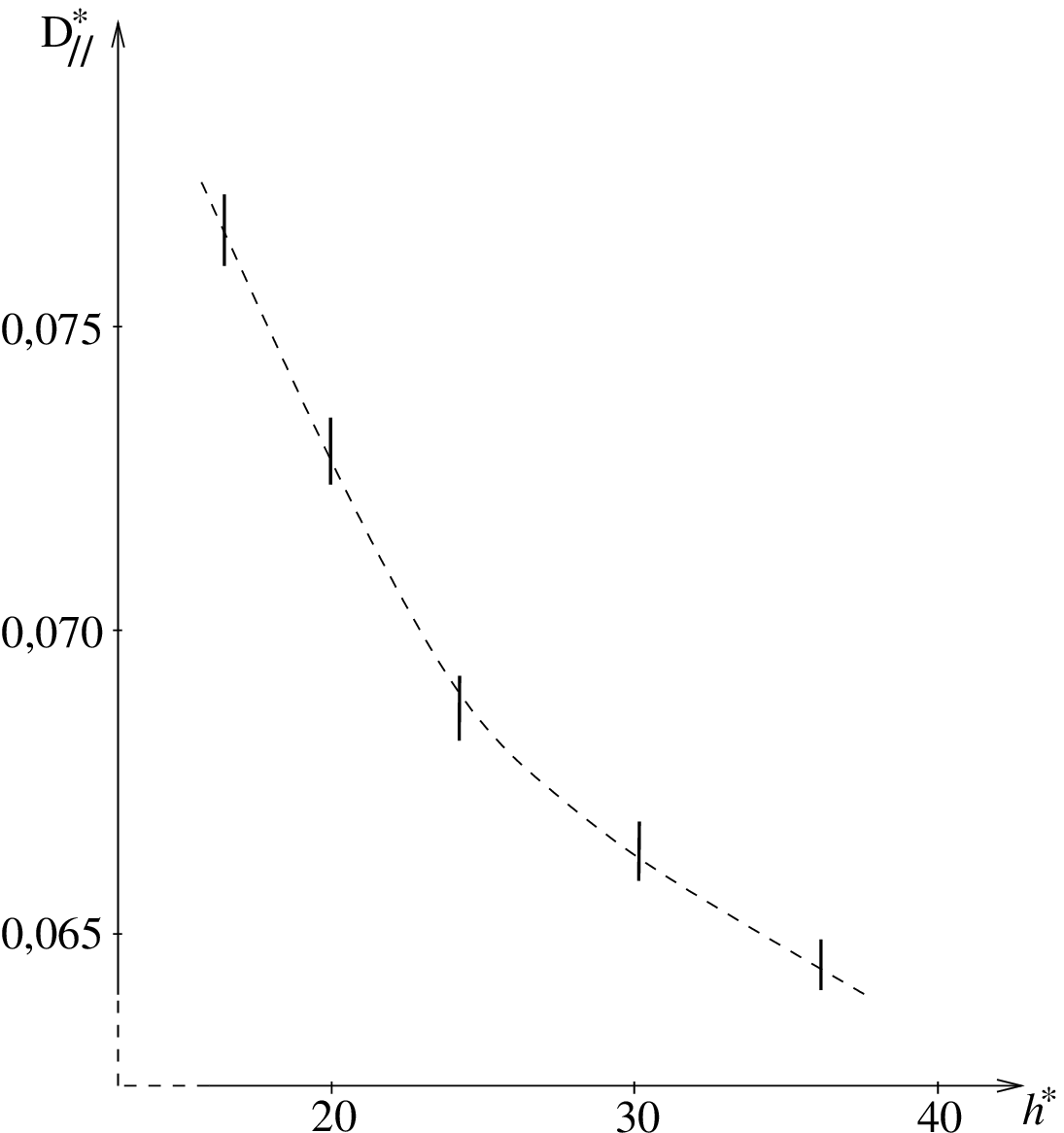,width=7cm}}}
\caption{\textit{Diffusion constant (in Lennard
Jones units) as a function of the pore width   $h^\star= h/\sigma$
for a Lennard Jones fluid confined between two parallel walls.
The density in the middle of the pore is $0.92 \sigma^{-3}$ for panel (a) and
 $0.86 \sigma^{-3}$ for panel (b). $\delta^*=\delta/\sigma$ below
the panels indicates the slipping length corresponding to each considered 
fluid-solid 
interface (measured independently). The reduced temperature is $T=1$.}}
\label{diffu_md}
\end{figure}

In order to confirm qualitatively the general trend 
predicted in the above section, we present in figure \ref{diffu_md}
the results for the diffusion constant in a Lennard Jones fluid
confined between two solid walls. 
The two cases correspond to a 
wetting situation with a zero slipping length, and to a 
nonwetting one with a large slipping length. We stress that
while varying the pore width $h$, some care has been taken to
keep the density of the fluid
at the {\it center of the pore} fixed to its ``bulk'' value,
so that variations of the diffusion constant can only originate
from confinment contributions.

In the nonwetting
case, the increase of the diffusion constant is clearly visible
as soon as the pore size becomes smaller than the slip length.
Each of these curves corresponds to a constant $\delta$ 
cut of the surface in figure \ref{3d}.

\section{Conclusion}
Both  hydrodynamic arguments and microscopic simulations
indicate that the diffusion of a tagged molecule in a confined geometry
will indirectly be correlated  to  the wetting properties
of the fluid, through the hydrodynamic boundary condition at the 
interface. In particular, an increase of the diffusion constant with
confinment is predicted in the "nonwetting" case. We emphasize, however, that 
the mechanism discussed in this paper is quite generic,
and does not consider the possibility of specific interactions with the substrate. Care should also be taken in measuring the diffusion constant
in the bulk and in the confined medium under similar thermodynamic conditions 
(pressure), in order to make  a sensible comparison.

\end{document}